\documentclass[aps,pra,superscriptaddress,showpacs,twocolumn]{revtex4}
\usepackage{graphicx}
\usepackage{amssymb}
\usepackage{amsmath}

\newcommand{\bk}{{\bf k}}

\newcommand{\br}{{\bf r}}

\newcommand{\kB}{k_{\mathrm{B}}}
\newcommand{\kF}{k_{\mathrm{F}}}
\newcommand{\nF}{n_{\mathrm{F}}}
\newcommand{\Green}{{G}}

\begin{document}

\title{Integral equation for inhomogeneous condensed bosons
   generalizing the Gross-Pitaevskii differential equation}

\author{G. G. N. Angilella}
\affiliation{Dipartimento di Fisica e Astronomia, Universit\`a di
   Catania,\\ and Istituto Nazionale per la Fisica della Materia,
   UdR di Catania,\\ Via S. Sofia, 64, I-95123 Catania, Italy}
\author{N. H. March}
\affiliation{Department of Physics, University of Antwerp,\\
Groenenborgerlaan 171, B-2020 Antwerp, Belgium}
\affiliation{Oxford University, Oxford, England} 
\author{R. Pucci}
\affiliation{Dipartimento di Fisica e Astronomia, Universit\`a di
   Catania,\\ and Istituto Nazionale per la Fisica della Materia,
   UdR di Catania,\\ Via S. Sofia, 64, I-95123 Catania, Italy}

\date{\today}

\begin{abstract}
\medskip
We give here the derivation of a Gross-Pitaevskii--type equation for
   inhomogeneous condensed bosons.
Instead of the original Gross-Pitaevskii differential equation, we
   obtain an integral equation that implies less restrictive
   assumptions than are made in the very recent study of Pieri and
   Strinati [Phys. Rev. Lett. {\bf 91}, 030401 (2003)].
In particular, the Thomas-Fermi approximation and the restriction to
   small spatial variations of the order parameter invoked in their
   study are avoided.
\\
\end{abstract} 

\maketitle

In a very stimulating recent contribution, Pieri and Strinati
   (referred to as PS below) \cite{Pieri:03} have `derived' the
   non-linear Gross-Pitaevskii differential equation for condensed
   bosons by taking 
   as their starting point the Bogoliubov-de~Gennes equation for
   superfluid fermions.

The purpose of this Brief Report is to demonstrate that one
   can generalize the zero-temperature differential
   Gross-Pitaevskii equation 
   while remaining within the original framework of PS, an
   integral equation formulation then resulting.
The framework of PS is provided by the coupled integral equations involving
   Green functions $\Green_{21}$, $\Green_{11}$ and $\tilde{\Green}_\circ$.
The equations are:
\begin{subequations}
\begin{eqnarray}
\Green_{11} (\br,\br^\prime ;\omega_s ) &=& \tilde{\Green}_\circ
   (\br,\br^\prime ;\omega_s ) + \int d\br^{\prime\prime} \tilde{\Green}_\circ
   (\br,\br^{\prime\prime} ;\omega_s ) \nonumber\\
&&\times \Delta(\br^{\prime\prime} )
   \Green_{21} ( \br^{\prime\prime} , \br^\prime ;\omega_s ),\\
\Green_{21} (\br,\br^\prime ;\omega_s ) &=& - \int d\br^{\prime\prime}
   \tilde{\Green}_\circ 
   (\br^{\prime\prime} ,\br ;-\omega_s ) \nonumber\\
&&\times \Delta^\ast (\br^{\prime\prime} )
   \Green_{11} ( \br^{\prime\prime} , \br^\prime ;\omega_s ),
\end{eqnarray}
\label{eq:PS}
\end{subequations}
\!\!\!\!\!
where $\omega_s = (2s+1)\pi/\beta$ ($s$ is an integer) is a fermionic
   Matsubara frequency, $\beta = 1/\kB T$, $\Green_{11}$ is the normal
   and $\Green_{21}$ 
   is the anomalous single-particle Green function.
The third Green function appearing in Eqs.~(\ref{eq:PS}), namely
   $\tilde{\Green}_\circ$, satisfies the equation
\begin{equation}
[i\omega_s -H(\br) ]\tilde{\Green}_\circ (\br,\br^\prime ; \omega_s ) = \delta
   (\br-\br^\prime ),
\label{eq:H}
\end{equation}
where the single-particle Hamiltonian $H(\br)$ is defined by:
\begin{equation}
H(\br) = - \frac{\nabla^2}{2m} + V(\br) -\mu,
\label{eq:H1}
\end{equation}
$\mu$ being the Fermionic chemical potential.
As PS stress, Eqs.~(\ref{eq:PS}), when taken together with the
   self-consistency equation for the gap function:
\begin{equation}
\Delta^\ast (\br) = \frac{V_0}{\beta} \sum_s \Green_{21}
   (\br,\br ; \omega_s ),
\label{eq:gap}
\end{equation}
are entirely equivalent to the Bogoliubov-de~Gennes equations that
   describe the behavior of superfluid fermions in the presence of an
   external potential.
Equations~(\ref{eq:PS}--\ref{eq:gap}) define what we have termed the
   original framework of the PS study.
The constant $V_0 <0$ entering Eq.~(\ref{eq:gap}) arises from the
   contact potential $V_0 \delta(\br -\br^\prime )$ assumed by PS to
   act between fermions with opposite spins.
We also retain here their use of the ratio $\Delta(\br)/\mu$ as an
   expansion parameter which allows the rapid truncation of such
   series, which then leads for strong coupling to an integral
   equation for the gap function
\begin{eqnarray}
-\frac{1}{V_0} \Delta^\ast (\br) &=& \int d\br_1 \, Q(\br,\br_1 )
   \Delta^\ast (\br_1 ) \nonumber\\
&&+\int d\br_1 d\br_2 d\br_3 \, R(\br,\br_1 ,\br_2 ,\br_3 )
\nonumber\\
&&\times
   \Delta^\ast (\br_1 ) \Delta(\br_2 ) \Delta^\ast (\br_3 ),
\label{eq:DDD}
\end{eqnarray}
where $R$ is written explicitly in terms of $\tilde{\Green}_\circ
   (\br,\br_1 ;\omega_s )$ in Eq.~(15) of PS.
However, as will emerge below, it is the non-local kernel
   $Q(\br,\br^\prime )$ which is at the heart of the present study.
In terms of the Green function $\tilde{\Green}_\circ$ entering
   Eq.~(\ref{eq:H}), $Q(\br,\br^\prime )$ is given by [PS: Eq.~(14)]:
\begin{equation}
Q(\br,\br^\prime ) = \frac{1}{\beta} \sum_s \tilde{\Green}_\circ
   (\br^\prime ,\br;-\omega_s ) \tilde{\Green}_\circ (\br^\prime ,\br
   ; \omega_s ).
\label{eq:PS14}
\end{equation}
We take the integral equation~(\ref{eq:DDD}) for the gap function as
   the starting point of this Brief Report.
For our purposes below, it is then crucial to gain insight into the
   kernel $Q$ in Eq.~(\ref{eq:PS14}), and in particular to carry out
   the summation explicitly over the Matsubara frequencies $\omega_s$.

To gain orientation, let us first perform this summation when the
   external potential $V(\br )$ is set to zero in Eq.~(\ref{eq:H}).
Having achieved this summation, we shall present a general method to
   allow the sum over $\omega_s$ to be achieved for $V(\br)\neq0$,
   using earlier work of Stoddart, Hilton and March
   \cite{Stoddart:68}.

\begin{figure}[t]
\centering
\includegraphics[height=0.9\columnwidth,angle=-90]{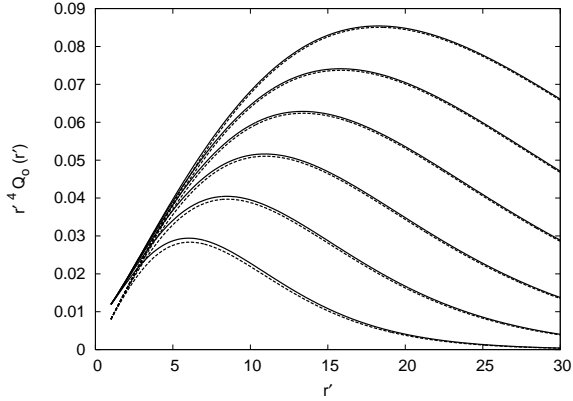}
\caption{Solid lines show $r^{\prime 4} Q_\circ (r^\prime )$, where $Q(r)$
   is defined by Eq.~(\protect\ref{eq:PS14}), as a
   function of $r^\prime = \kF r$, for
   several temperatures, given by $\beta^\prime = \beta\mu=10-30$
   (bottom to top).
Dashed lines are the asymptotic expansion Eq.~(\ref{eq:asymptotic}).
Units are such that $\kF^6/(4\mu) = 1$.}
\label{fig:Q0}
\end{figure}

Returning to the explicit form of $Q(\br,\br_1 )$ given in
   Eq.~(\ref{eq:PS14}) above, it is natural to study first the
   translational invariant, free-electron limit of
   Eq.~(\ref{eq:PS14}), say $Q_\circ (r)$, with 
   $r=|\br-\br_1 |$, which is obtained by `switching off' the one-body
   potential $V(\br)$.
This amounts to replacing $\tilde{G}_\circ$ in Eq.~(\ref{eq:PS14}) with
   the free-electron Green function $G_\circ$.
For the Fourier transform of $Q_\circ (r)$, we formally find
\begin{equation}
\hat{Q}_\circ (k) = \int \frac{d \bk^\prime}{(2\pi)^3}
\frac{1-\nF (\xi_{\bk-\bk^\prime} ) - \nF (\xi_{\bk^\prime}
   )}{\xi_{\bk-\bk^\prime} + \xi_{\bk^\prime}} ,
\label{eq:QF0}
\end{equation}
where $\xi_\bk = k^2 /2m - \mu$ and $\nF (\xi)$ is the Fermi-Dirac
   distribution function.
However, it should be noted that, in three dimensions,
   Eq.~(\ref{eq:QF0}) contains a divergent contribution at large
   wave-numbers, which implies a divergent behavior of $Q_\circ (r)$
   at small distances $r$.
Indeed, we find the asymptotic expansion (see also
   Ref.~\cite{Alexandrov:03}):
\begin{equation}
\frac{4\mu}{\kF^6}
Q_\circ (r) 
\sim \frac{1}{4\pi^2} \frac{1}{r^{\prime 2} \beta^\prime } \frac{1}{\sinh a},
\quad \beta^\prime \gg 1,
\label{eq:asymptotic}
\end{equation}
where $r^\prime = \kF r$, $\kF$ is the Fermi wave-number, defined by
   $\mu=\kF^2 /2m$, $\beta^\prime = \beta\mu$, and $a=r^\prime
   \pi/\beta^\prime$. 
Fig.~\ref{fig:Q0} shows then our numerical results for $r^{\prime 4} Q_\circ
   (r^\prime )$, as a function of $r^\prime $, for several
   temperatures ($\beta^\prime =10-30$).

Following Stoddart \emph{et al.} \cite{Stoddart:68}, the canonical
   density matrix $C(\br,\br^\prime ,\beta)$ is defined by
\begin{equation}
C(\br,\br^\prime ,\beta) = \sum_i \psi_i (\br) \psi_i^\ast (\br^\prime )
   e^{-\beta\epsilon_i} ,
\end{equation}
where $\beta = 1/\kB T$.
Within the perturbative approach of March and Murray
   \cite{March:60,March:61}, with plane waves as the unperturbed
   solution, the canonical density matrix can then be written to all
   orders in the external potential $V(\br)$ in terms of the
   free-particle canonical density matrix given by 
\begin{equation}
C_0 (z,\beta) = (2\pi\beta)^{-3/2} \exp (-z^2 /2\beta ),
\label{eq:C0}
\end{equation}
as
\begin{equation}
C(\br,\br_0 ,\beta) = \int_0^\infty d z\, z\, C_0 (z,\beta)
   f(z,\br,\br_0 ),
\label{eq:C}
\end{equation}
where $f$ satisfies the integral equation \cite{Stoddart:68}:
\begin{eqnarray}
f(z,\br,\br_0 ) &=& \frac{1}{z} \delta (z-|\br-\br_0 |) 
- \int d\br_1 \frac{V(\br_1 )}{2\pi |\br-\br_1 |} \nonumber\\
&&\times f(z-|\br-\br_1 |,\br_1 ,\br_0 ).
\label{eq:f}
\end{eqnarray}
The desired Green function $\tilde{\Green}_\circ$ is then to be
   obtained from $f$ entering Eqs.~(\ref{eq:C}) and (\ref{eq:f}) as
   \cite{Stoddart:68}
\begin{equation}
\tilde{\Green}_\circ (\br,\br_1 ; k ) = \int_0^\infty d z \, z\,
\bar{\Green}_\circ (z; k) f(z,\br,\br_1 ),
\label{eq:SG}
\end{equation}
where
\begin{equation}
\bar{\Green}_\circ (z;k) = \frac{e^{ikz}}{4\pi z} .
\end{equation}
One may also take advantage of the expression in Eq.~(\ref{eq:SG}) of
   $\tilde{\Green}_\circ$ in terms of $\bar{\Green}_\circ$ to rewrite the kernel
   $Q(\br,\br_1 )$ defined by Eq.~(\ref{eq:PS14}) as
\begin{equation}
Q(\br,\br_1 ) = \int_0^\infty dz_1 \, dz_2 \, z_1 \, z_2 \, 
f(z_1 ,\br_1 ,\br) f(z_2 ,\br_1 ,\br) Q_\circ (z_1 ,z_2 ),
\label{eq:Qff}
\end{equation}
where the Fourier transform of $Q_\circ (z_1 ,z_2 )$ is given by
\begin{equation}
\hat{Q}_\circ (\bk_1 ,\bk_2 ) = \frac{1-\nF(\xi_{\bk_1} ) - \nF(\xi_{\bk_2}
   )}{\xi_{\bk_1} + \xi_{\bk_2}} .
\end{equation}
Hence, the sum over Matsubara frequencies has still been carried out
   in the presence of an external potential $V(\br)$ entering
   Eq.~(\ref{eq:f}) for the function $f$.

\begin{figure}[t]
\centering
\includegraphics[height=0.9\columnwidth,angle=-90]{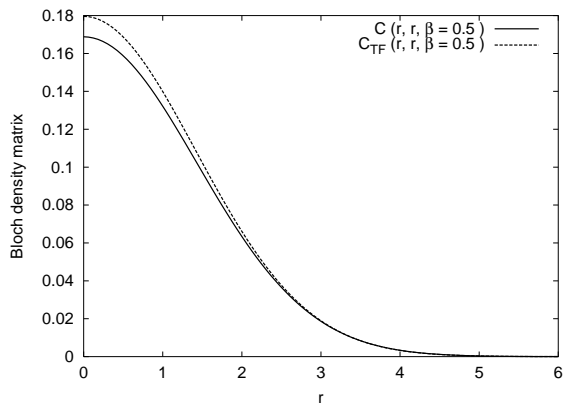}
\caption{Shows diagonal element of the canonical density matrix
   $C(\br,\br,\beta)$, Eq.~(\protect\ref{eq:harmonicC}), and its
   Thomas-Fermi approximation, Eq.~(\protect\ref{eq:harmonicCTF}), as
   a function of $r$, for $\beta=0.5$.
Energies are in units of $\hbar\omega$, while lengths are in units of
   $(\hbar/m\omega)^{1/2}$.}
\label{fig:C}
\end{figure}

Because of current interest in harmonic confinement in magnetic traps
   at low temperatures, let us illustrate the rather formal
   Eqs.~(\ref{eq:C}) and (\ref{eq:f}) when the external potential
   $V(\br)$ has the explicit isotropic harmonic oscillator form in
   three dimensions, namely
\begin{equation}
V(\br) = \frac{1}{2} m\omega^2 r^2 .
\label{eq:harmonic}
\end{equation}
Following the pioneering work of Sondheimer and Wilson
   \cite{Sondheimer:51} on free electrons in a magnetic field, the
   diagonal element $C(\br,\br,\beta)$ when $V(\br)$ is given by
   Eq.~(\ref{eq:harmonic}) takes the form (see \emph{e.g.}
   \cite{March:95}, p.~27; see also \cite{Howard:03})
\begin{eqnarray}
C(\br,\br,\beta) &=& \left( \frac{m}{2\pi\hbar} \right)^{3/2} \left(
   \frac{\omega}{\sinh \hbar\omega\beta} \right)^{3/2} \nonumber\\
&&\times \exp \left( -\frac{m}{\hbar} \omega r^2 \tanh \frac{1}{2}
   \hbar\omega\beta \right),
\label{eq:harmonicC}
\end{eqnarray}
which is the so-called Slater sum of quantum chemistry (Fig.~\ref{fig:C}).

From Eqs.~(\ref{eq:C0}) and (\ref{eq:C}), performing the substitution
   $t=z^2 /2$, it then follows that 
   $f(z,\br,\br_0 )$ can be expressed as the inverse Laplace transform
\begin{equation}
f(z,\br,\br_0 ) = (2\pi)^{3/2} \mathcal{L}^{-1} \left[
s^{-3/2} C(\br,\br_0 ,s^{-1} ) \right]
\label{eq:laplace}
\end{equation}
where $(t,s)$ are conjugate variables with respect to the Laplace
   transform.

Within the Thomas-Fermi (TF) approximation, we take:
\begin{equation}
C_{\mathrm{TF}} (\br,\br,\beta) = \frac{1}{(2\pi\beta)^{3/2}} \exp
   [-\beta V(\br) ],
\label{eq:harmonicCTF}
\end{equation}
which is plotted also in Fig.~\ref{fig:C} for $V(\br)$ given by
   Eq.~(\ref{eq:harmonic}).
For the value of $\beta$ shown, the TF form Eq.~(\ref{eq:harmonicCTF})
   is seen to be a useful approximation to the exact result,
   Eq.~(\ref{eq:harmonicC}).
Inserting Eq.~(\ref{eq:harmonicCTF}) into Eq.~(\ref{eq:laplace}) we
   find
\begin{equation}
f_{\mathrm{TF}} (z,\br,\br) = \frac{\delta(z)}{z} - \frac{\sqrt{2V(\br)}}{z} J_1
   [\sqrt{2V(\br)} z ],
\label{eq:fTF}
\end{equation}
where $J_1$ denotes the Bessel function of the first kind and order
   one.

\begin{figure}[t]
\centering
\includegraphics[height=0.9\columnwidth,angle=-90]{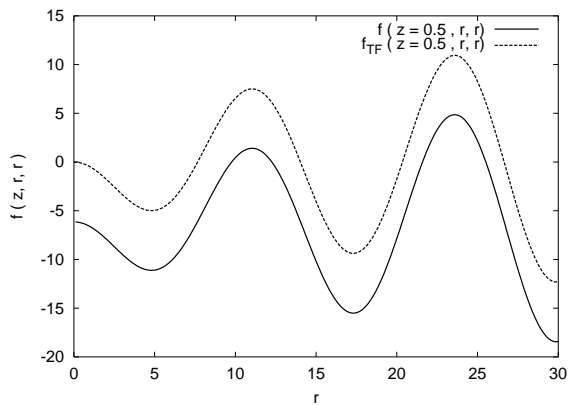}
\caption{Shows diagonal $f(z,\br,\br)$ corresponding to the harmonic
   potential, as given by the inverse Laplace transform,
   Eq.~(\protect\ref{eq:laplace}), as well as the regular part of its
   Thomas-Fermi approximation, Eq.~(\ref{eq:fTF}), as a function of
   $r$, for fixed $z=0.5$.
Units as in Fig.~\protect\ref{fig:C}.}
\label{fig:f}
\end{figure}

Fig.~\ref{fig:f} shows $f(z,\br,\br)$ as a function of $r$ for fixed
   $z$, as obtained by numerically performing the inverse Laplace
   transform in Eq.~(\ref{eq:laplace}) for the harmonic potential
   case.
The regular contribution to the analytic result for the Thomas-Fermi
   approximation, Eq.~(\ref{eq:fTF}), is also plotted for comparison.
The similarity in shape between approximate and (numerically) exact
   results for this harmonic confinement model seems to us rather
   remarkable.
After this model test of a TF-like approximation invoked by PS, we
   return to the general case, based on the exact result
   Eq.~(\ref{eq:Qff}) for the kernel $Q(\br,\br_1 )$.

Then, we invert the argument of PS but still use a further essential
   assumption of their study, namely that the condensate wave function
   $\Phi(\br)$ entering the Gross-Pitaevskii equation is related to
   the gap function $\Delta(\br)$ by
\begin{equation}
\Phi(\br) = \left( \frac{m^2 a_{\mathrm{F}}}{8\pi} \right)^{1/2}
   \Delta(\br) \equiv k \Delta(\br).
\label{eq:k}
\end{equation}
Here, in the strong coupling limit, and following PS, $a_{\mathrm{F}}
   \sim (2m|\mu|)^{-1/2}$ represents the characteristic length scale
   for the non-interacting Green function $\Green_\circ$, equal to
   $\tilde{\Green}_\circ$ above when $V(\br)$ is put equal to zero.

Given the validity of this PS assumption, Eq.~(\ref{eq:k}), we then
   rewrite Eq.~(\ref{eq:DDD}) as an equation for $\Phi(\br)$:
\begin{eqnarray}
-\frac{1}{V_0} \Phi^\ast (\br) &=& \int d\br_1 \, Q(\br,\br_1 )
   \Phi^\ast (\br_1 ) \nonumber\\
&&+\frac{1}{k^2} 
\int d\br_1 d\br_2 d\br_3 \, R(\br,\br_1 ,\br_2 ,\br_3 )
\nonumber\\
&&\times
   \Phi^\ast (\br_1 ) \Phi(\br_2 ) \Phi^\ast (\br_3 ).
\label{eq:PPP}
\end{eqnarray}
This then is the proposed generalization of the Gross-Pitaevskii
   equation, but with $Q(\br,\br_1 )$ to be calculated more
   accurately than by the Thomas-Fermi--like assumption of Pieri and
   Strinati \cite{Pieri:03}, via Eqs.~(\ref{eq:Qff}) and (\ref{eq:f}).

While Eq.~(\ref{eq:PPP}) is a direct consequence of the above
   arguments, it remains an expansion in $\Phi$, in suitable reduced
   form.
Therefore, a first attempt to simplify this Eq.~(\ref{eq:PPP}) is to retain the
   approximation given by the Pieri-Strinati approach in the
   `smallest' term involving $O(\Phi^3 )$ on the right-hand side of
   the basic Eq.~(\ref{eq:PPP}).
Thus one reaches the (still non-local) equation for the condensate
   wave function $\Phi(\br)$:
\begin{eqnarray}
-\frac{1}{V_0} \Phi^\ast (\br) &=& \int d\br_1 \, Q(\br,\br_1 )
   \Phi (\br_1 ) \nonumber\\
&&-\frac{ma_{\mathrm{F}}^2}{2} |\Phi(\br)|^2 \Phi(\br).
\label{eq:PPPP}
\end{eqnarray}

For sufficiently small spatial variations in the condensate wave
   function $\Phi(\br)$ in Eq.~(\ref{eq:PPPP}), the basic nonlocality
   can be removed by Taylor expanding $\Phi(\br_1 )$ around the
   position $\br$ in the integral term.
This then characterizes the problem in terms of `partial moments' of
   the kernel $Q(\br,\br_1 )$, namely $\smallint Q(\br,\br_1 )d\br_1$
   and $\smallint Q(\br,\br_1 )|\br-\br_1 |^2 d\br_1$.
Such partial moments then enter the original Gross-Pitaevskii
   equation, as stressed by PS.

In summary, we propose the retention of the non-local kernel
   $Q(\br,\br_1 )$ as in Eq.~(\ref{eq:PPP}) above, since the sum over
   Matsubara frequencies in Eq.~(\ref{eq:PS14}) has been performed in
   Eq.~(\ref{eq:Qff}), which is a central result of the present study.
However, in the terms of $O(\Phi^3 )$ in Eq.~(\ref{eq:PPP}), a
   sensible starting point is to follow the PS approximation displayed
   in Eq.~(\ref{eq:PPPP}).

As to future directions, evaluation of the non-local kernel in
   Eq.~(\ref{eq:Qff}) for other external potentials than the harmonic
   case in Eq.~(\ref{eq:harmonic}) is of obvious interest.
For this latter model, though our Fig.~\ref{fig:f} considers the
   diagonal element of $f(z,\br ,\br_1 )$, the off-diagonal form of
   $C(\br,\br_1 ,\beta)$ is known \cite{Howard:03}, and numerical
   Laplace inversion to obtain $f(z,\br ,\br_1 )$ is entirely feasible.
Then $Q(\br,\br_1 )$ can be obtained, though of course numerically.

The Gross-Pitaevskii equation is valid in the strong-coupling limit of
   superfluidity.
It has to be stressed that in the weak coupling limit one can
   also derive a Ginzburg-Landau equation starting from the
   Bogoliubov-de~Gennes equations.
We note specifically in this context that the derivation of the
   Ginzburg-Landau equation in the weak-coupling limit for the
   harmonic trap was presented by Baranov and Petrov
   \cite{Baranov:98}.
The results presented in this Brief Report
   are also relevant to the weak-coupling limit of
   superfluidity.

Finally, we should mention the very recent discussions of the
   foundations of the Gross-Pitaevskii equation by Leggett
   \cite{Leggett:03}.
He concludes that there is no correlated many-body wave-function
   underlying their original equation.
It will be interesting for the future to know whether the non-local
   versions of Eqs.~(\ref{eq:PPP}) and (\ref{eq:PPPP}) proposed here
   are still subject to this limitation.

\begin{acknowledgments}
G.G.N.A. thanks N. Andrenacci and F. S. Cataliotti for
   useful discussions, and K. V. Kamenev for helpful correspondence.
N.H.M. brought his contribution to the present study to fruition
   during a visit to Catania in 2003.
He thanks the Department of Physics and Astronomy for the stimulating
   atmosphere and for generous hospitality.
\end{acknowledgments}

\begin{small}
\bibliographystyle{apsrev}
\bibliography{a,b,c,d,e,f,g,h,i,j,k,l,m,n,o,p,q,r,s,t,u,v,w,x,y,z,zzproceedings,Angilella}
\end{small}

\end{document}